\documentclass[sn-mathphys-num]{sn-jnl}

\usepackage{graphicx}%
\usepackage{multirow}%
\usepackage{amsmath,amssymb,amsfonts}%
\usepackage{amsthm}%
\usepackage{mathrsfs}%
\usepackage[title]{appendix}%
\usepackage{xcolor}%
\usepackage{textcomp}%
\usepackage{manyfoot}%
\usepackage{booktabs}%
\usepackage{algorithm}%
\usepackage{algorithmic}%
\usepackage{listings}%
\usepackage{tabularx}

\theoremstyle{thmstyleone}%
\theoremstyle{thmstyletwo}%

\theoremstyle{thmstylethree}%

\begin{document}

\title[Article Title]{Garbage in Garbage out: Impacts of data quality on criminal network intervention}


\author[1, 3]{\fnm{Wang Ngai} \sur{Yeung}}

\author[1,4]{\fnm{Riccardo} \sur{Di Clemente}}

\author*[2]{\fnm{Renaud} \sur{Lambiotte}}\email{renaud.lambiotte@maths.ox.ac.uk}

\affil[1]{\orgdiv{Complex Connections Lab, Network Science Institute}, \orgname{Northeastern University London}, \orgaddress{\street{Devon House}, \city{London}, \postcode{E1W 1LP}, \country{United Kingdom}}}

\affil[2]{\orgdiv{Mathematical Institute}, \orgname{University of Oxford}, \orgaddress{\street{Andrew Wiles Building}, \city{Oxford}, \postcode{OX2 6GG}, \country{United Kingdom}}}

\affil[3]{\orgdiv{Oxford Internet Institute}, \orgname{University of Oxford}, \orgaddress{\street{1 St Giles}, \city{Oxford}, \postcode{OX1 3JS}, \country{United Kingdom}}}

\affil[4]{\orgname{ISI Foundation}, \orgaddress{\street{Via Chisola 5}, \city{Turin}, \postcode{10126}, \country{Italy}}}


\abstract{Criminal networks such as human trafficking rings are threats to the rule of law, democracy and public safety in our global society. Network science  provides invaluable tools to identify key players and design interventions for Law Enforcement Agencies (LEAs), e.g., to dismantle their organisation. However, poor data quality and the adaptiveness of criminal networks through self-organization make effective disruption   extremely challenging. Although there exists a large body of work building and applying network scientific tools to attack criminal networks, these work often implicitly assume that the network measurements are  accurate and complete. Moreover, there is thus far no comprehensive understanding of the impacts of data quality on the downstream effectiveness of interventions. This work investigates the relationship between data quality and intervention effectiveness based on classical graph theoretic and machine learning-based approaches. Decentralization emerges as a major factor in network robustness, particularly under conditions of incomplete data, which renders attack strategies largely ineffective. Moreover, the robustness of centralized networks can be boosted using simple heuristics, making targeted attack more infeasible. Consequently, we advocate for a more cautious application of network science in disrupting criminal networks, the continuous development of an interoperable intelligence ecosystem, and the creation of novel network inference techniques to address data quality challenges.}

\keywords{Criminal network, Data quality, Complex system, Organized crime}



\maketitle

\section{Introduction}\label{sec1}
Criminal organizations are ubiquitous and the \textit{Dark Networks} that support their operations  are threats to our democracy, the rule of law and public safety \cite{Raab2003}.  Criminal networks  operate outside of the law in various contexts, such as drug trafficking rings \cite{Bright2011} and terrorist organizations \cite{Xu2008}. For example, in 2008, 2.3\% of the Australian population whose age are over 14 had consumed methamphetamine within 12 months \cite{Bright2011}, revealing an underlying public health issue across the country. Terrorist organizations, such as the Global Salafi Jihad (GSJ) network, which includes al-Qaeda and was responsible for large-scale attacks like 9/11, are among the more extensively studied criminal networks. More recently, within the European Union (EU), around thirty transnational criminal networks are active across most member countries, driving violent and exploitative crimes such as burglary and sex trafficking \cite{european_union_agency_for_law_enforcement_cooperation_decoding_2024}.

 Although criminal networks are widely recognized and monitored by governments worldwide, disrupting them remains a significant challenge for Law Enforcement Agencies (LEAs) and intelligence agencies. One key obstacle is the advancement of secure communication technologies, which enable criminal organizations to coordinate illegal activities with greater efficiency and reduced detection \cite{Crossley2012}. The persistence and resilience of these organizations are often reinforced by their adaptability, reliance on corruption, and use of forensic countermeasures, such as encrypted devices like SkyECC, to evade monitoring. Over the past two decades, these covert networks have also become increasingly decentralized \cite{Cui2024}. Additionally, the collection, management, and interpretation of criminal data are frequently flawed. For large networks, such as the Sicilian Mafia or outlaw motorcycle gangs (OMCGs), surveillance often fails to capture critical communications, resulting in substantial data gaps \cite{Baker1993}. Conversely, in smaller networks, intelligence gathered from reports or investigations may be overlooked or lost due to corruption or poor judgment \cite{Morselli2009}. As a result, data on covert networks is often incomplete, inaccurate, and unreliable. Finally, the diversity of network topologies and dynamics across various types of covert organizations complicates efforts to develop a unified approach to network disruption.

These significant challenges lie at the core of contemporary research on covert networks and their disruption. Over the past two decades, network science has emerged as a critical tool for intervening in criminal networks. By leveraging new data sources, such as cellphone call records, network science has been applied to develop interactive strategies that assist in suspect identification and in revealing the hidden structures of criminal organizations \cite{Zhou2017}. A key approach in these studies is quantifying individual importance through centrality measures. Indeed, a substantial body of literature on criminal network disruption evaluates the most effective node-ranking strategies for targeting or apprehending actors within these networks \cite{DelaMoraTostado2024,Cavallaro2020,Duijn2014,Holme2002}. However, a primary limitation of these studies is their reliance on the assumption of data completeness and accuracy, which can lead to overfitting on potentially flawed data. Not only is data collection extremely challenging in this context \cite{Duijn2015}, but criminal networks can also manipulate their structures—using tactics such as the Remove-One-Attach-Many (ROAM) heuristic to obscure leaders and captains \cite{waniek_hiding_2018} or optimizing network design to create nodes with identical centralities \cite{Dey_2019}.

Overall, findings from previous studies on the effectiveness of attack strategies should be interpreted with careful consideration of issues related to missing and inaccurate data. While recent work has begun addressing the impact of missing data in criminal networks (see \cite{DeMoor2020}), these studies have largely concentrated on (1) large datasets and (2) network estimation errors. In this context, the present work examines the core question of how data quality—whether compromised by incompleteness, inaccuracy, or intentional network self-alteration—affects the effectiveness of downstream network interventions in smaller networks. Through a series of numerical experiments, our results indicate that missing data renders most node-ranking methods ineffective at reducing the Largest Connected Component (LCC) in both centralized and decentralized networks. This limitation is further exacerbated by the potential for network topology to be restructured through simple heuristics. Based on these findings, we advocate for heightened awareness of the limitations of network science in disrupting criminal networks under various scenarios of poor data quality and emphasize the need for ongoing advancements in data collection and annotation methods.

\section{\label{results}Results}

While numerous robustness measures exist \cite{Schwarze2024}, we choose sequential node percolation for our analysis due to its two primary advantages. First, it is widely used as a method to study network vulnerability. Second, node percolation serves as an abstract representation of real-life interventions in criminal networks, such as the arrest and incarceration of key individuals. For this purpose, we selected various node-ranking methods based on classical centrality measures, as well as more advanced heuristics-based and machine learning-based approaches (see details in \autoref{method}). We then used these ranking strategies to conduct percolation experiments on the networks.

We use publicly available static network data, specifically: (1) the London juvenile gang network \cite{grund_ethnic_2015}, (2) the `Ndrangheta network \cite{Legramanti2022}, (3) the New York cocaine trafficking ring \cite{Natarajan2006}, and (4) the Madrid train bombing terrorist networks \cite{Walther2014}. These networks were selected to provide diversity in size, topology, and organizational goals, allowing for a comparative examination of intervention effectiveness (see \autoref{tab:data-description}). Although the data sources occasionally include demographic variables and edge attributes (e.g., relationships between actors), our analysis treats these networks as unattributed, unweighted, and undirected graphs to facilitate more granular manipulation of network topologies. Building on experiments with the baseline networks (i.e., networks as represented in the original datasets), we conducted simulations to assess intervention effectiveness on perturbed networks (i.e., networks with missing, inaccurate, or topologically altered data).

\subsection{\label{sec:node_sim}Baseline performance of network intervention}

\begin{figure}[ht]
    \centering
    \includegraphics[width=\linewidth]{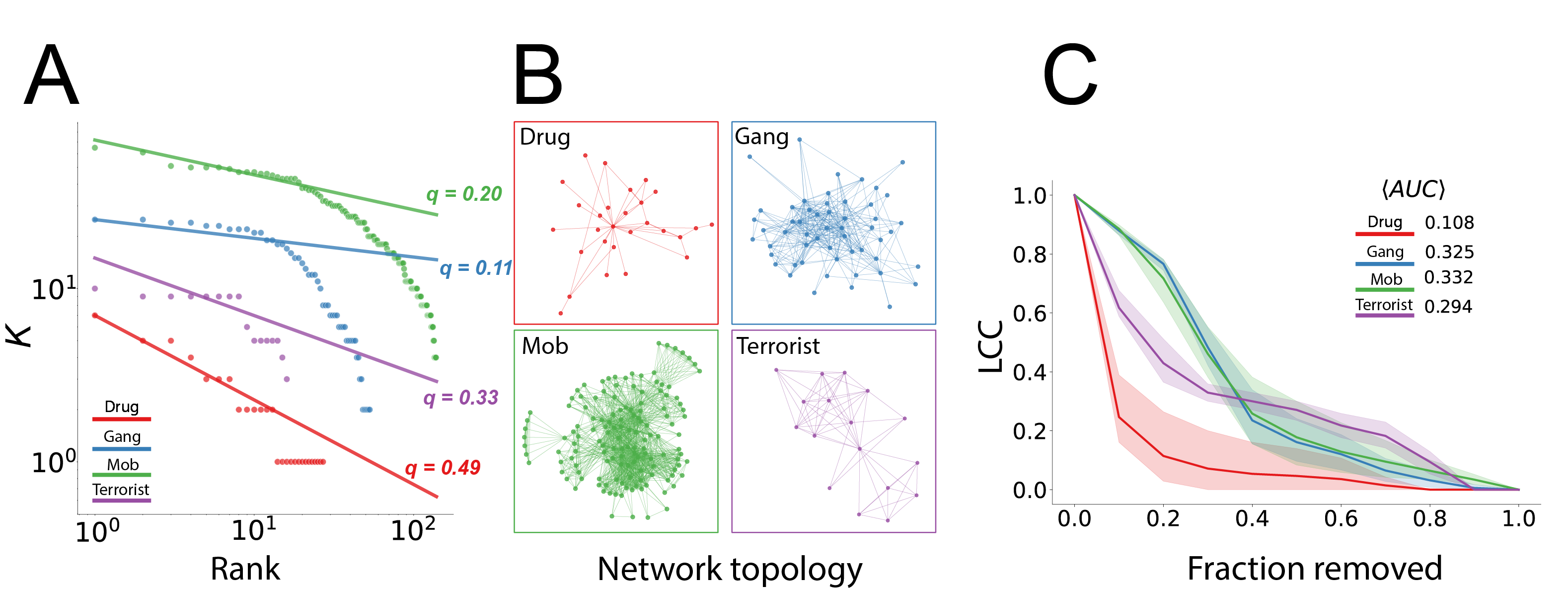}
    \caption{Criminal networks investigated in this work. \textbf{A}: Rank-Degree distribution fitted with Rank-Size scaling law $P_k = P_1 k^{-q}$, where $P_1$ denotes the highest degree and $k \in [1, 2, \cdots, N]$ refers to the rank. Rank is normalized for comparative purposes. Note that not all networks were well-fitted due to varying levels of network centralization. \textbf{B}: Visualization of the networks. \textbf{C}: Baseline percolation with shaded area is the standard deviation $\sigma$ and $\langle AUC \rangle$ is the average Area Under the Curve (AUC) of the LCC trajectories across all node-ranking methods.}
    \label{fig:fig1}
\end{figure}

In \autoref{fig:fig1}, we report the baseline performance of criminal network intervention. We quantify the network intervention effectiveness by approximating Area Under the Curve (AUC) of the size of Largest Connected Component (LCC) using the Trapezoidal Rule

\begin{equation}
    \text{AUC} = \int_a^b f(x)dx \approx \sum_{i=0}^{n-1} h \frac{f(x_i) + f(x_i+1)}{2}.
\end{equation}
The lower the AUC, the more effective an intervention is. We  observe that there exists a generally wide standard deviation of effectiveness $\sigma$ across all empirical datasets, but notably more so in the earlier phase of the New York cocaine trafficking ring and in most parts in the `Ndrangheta network. However, the average AUC is much lower for the cocaine trafficking network ($\langle AUC \rangle = 0.108$) than the mob network ($\langle AUC \rangle$ = 0.332), indicating the resilient nature of the mob network.

Another intriguing property is network centralization. As mentioned earlier, decentralization has been shown to be a significant factor for network resilience. To quantify centralization, we used Freeman degree centralization coefficient ($C_d$) \cite{Freeman1978}

\begin{equation}
    C_d = \frac{\sum_{v} \max_w c_w - c_v}{n^2 - 3n + 2},
\end{equation}
where $i$ denotes the node with highest degree centrality. The denominator $n^2 - 3n +2$ is the theoretical maximum sum of difference in degree given that a graph with one dominant node (e.g., a star graph) must have a degree of $n-1$ if self-loops are prohibited. For the other nodes in the graph, then, the degree will automatically be 1 and thus the difference between the dominant node and any follower node is  $(n-1) - 1 = n-2$. Altogether, the maximum sum of difference for all $n-1$ dominant-follower pairs is $(n-2)(n-1) = n^2 - 3n +2$. 

Our nuerical experiments indicate that networks with lower degree centralization—i.e., decentralized networks—generally exhibit greater resilience to network interventions. Notably, degree centralization is closely correlated with the rank-degree distribution. Networks with a rank-degree distribution that closely follows a typical Zipf scaling law, such as the Cocaine trafficking ring ($q = 1.20$) and the Madrid bombing terrorist network ($q = 0.38$), showed lower average AUC values compared to the more resilient, decentralized networks. It is worthwhile to mention that the decentralized networks also exhibits weaker power-law scaling, with $q = 0.27$ and $q = 0.32$ for the gang and mob networks respectively. This result was also consistent in synthetic networks with matching centralization configurations.

\subsection{\label{sec:missing}Influence of missing data on disruption effectiveness}

In this section, we examine the impact of missing data on network disruption effectiveness. Building on the baseline established earlier, we repeated the percolation experiments using randomly sampled subgraphs of the original graphs, assuming these graphs represent the true underlying network structure of the target organization. To simulate data incompleteness, subgraphs were generated by selecting a random fraction
$q$ of nodes from the underlying graph. Each simulation was conducted 
$10^3$ times per node-ranking method across all data completeness scenarios.

\begin{figure}[]
    \centering
    \includegraphics[width=\linewidth]{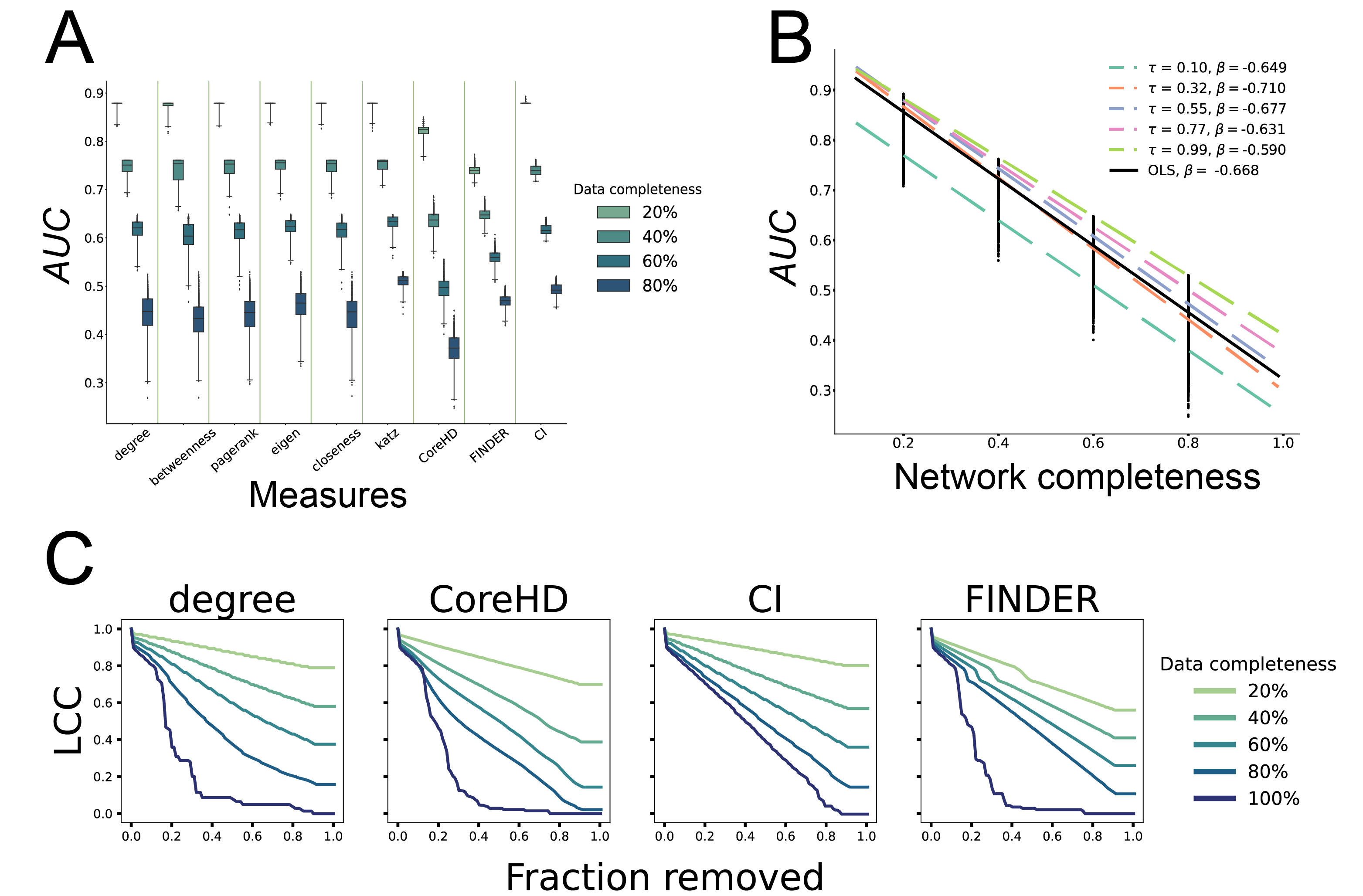}
    \caption{Impact of data incompleteness on percolation effectiveness in the `Ndrangheta network with $10^3$ simulations per node-ranking method across all data incompleteness scenarios. \textbf{A}: Boxplots of all node-ranking methods under different data completeness scenarios. Whiskers show the inter-quantile range (IQR) of the AUC and the outliers are indicated by small dots. \textbf{B}: Quantile regression of the effect of data completeness on AUC. \textbf{C}: Percolation plots of four different node-ranking methods under different data scenarios averaged over simulations.}
    \label{fig:missing_data_empirical}
\end{figure}

We observed that data incompleteness affected different networks similarly, regardless of the extent of incompleteness. For instance, although the New York cocaine trafficking ring and the London gang network had markedly different baseline performance, AUC values increased almost linearly as data completeness decreased. Thus, data incompleteness poses a challenge for the attacker irrespective of network topology, although centralized networks—like the New York cocaine trafficking ring and the Madrid bombing terrorist network—were still found to be more vulnerable to attacks under missing data conditions.
\autoref{fig:missing_data_empirical} illustrates the specific effects of data incompleteness on the Ndrangheta network. In general, for more resilient networks, such as the London gang and Ndrangheta networks, data incompleteness proves highly problematic for the attacker, as percolation effectiveness remains weak even with a relatively low percentage of missing data when using previously effective methods like FINDER and CI (see \hyperref[fig
]{Figure 2C}). In other words, to dismantle these networks effectively, the observer would need near-complete knowledge of the network structure.

We conducted additional quantile regression analysis with network completeness $q$ as the independent variable and attack efficiency (AUC) as the outcome variable to further confirm their relationship. For instance, on the `Ndrangheta network, as illustrated in \hyperref[fig:missing_data_empirical]{Figure 2B}, we observed a significant negative effect of network completeness on the value of AUC ($p < .001, t = -1672.5$) with a 95\% confidence interval of $[-0.686, -0.685]$. This effect, however, varied across quantiles  ($\tau$) of the outcome distribution. Specifically, we detected a non-linear association between $\tau$ and the regression coefficient $\beta_{\text{q}}$, showing the strongest negative effect near the median of the efficiency distribution, a weaker effect at the lower end of efficiency (where AUC is high), and the weakest effect at the higher end (where AUC is low).

Similar results were observed in the synthetic networks. Additionally, we found that ER and WS graphs exhibit high robustness against attacks, particularly in scenarios with significant data incompleteness. This resilience makes ER and WS graphs especially challenging to attack when data incompleteness is present.

\subsection{\label{sec:incomplete}Influence of inaccurate data on disruption effectiveness}

We present here the impact of data inaccuracy on disruption effectiveness. Unlike data incompleteness, where nodes may be missing, data inaccuracy retains the same number of observed nodes ($q = 1$) but introduces modifications to the edges. We define the inaccuracy rate as $p^k \in [0,1]$, where $k$ represents the number of incorrectly captured edges per node. This inaccuracy is modeled by adding, deleting, or non-degree-preserving rewiring of edges between actors. Note that for any node with a number of edge(s) smaller than $k$, we do edge modification for all of its edges. The results indicate only marginal differences in disruption effectiveness for graphs with small perturbations through such edge modifications. 

In the conservative scenario where only one inaccurate link ($k = 1$) can be modified for each node in their ego network, data inaccuracies exerted minimal negative impact on percolation effectiveness. Although a higher level of inaccuracy did delay the collapse of the largest connected component (LCC), shifting the trajectories rightward, this effect was consistent across all types of edge modifications. The influence of different types of edge modification on percolation effectiveness remained qualitatively similar. Consistent with our earlier findings, inaccuracies produced comparable effects across graphs with varying robustness levels. Our comparative analysis of ER and BAHK graphs under data inaccuracies revealed a relatively uniform delay in LCC dismantling with increasing inaccuracy. However, when we consider the less ideal case where multiple edges may be added to each inaccurate node, distinct patterns emerged. For instance, the less robust New York cocaine trafficking network experienced a more pronounced robustness shift due to inaccuracies than the London gang network. Despite general delays in both networks, the robustness shift for the New York network was significantly greater  ($\Delta AUC_{p^2 = 0} \approx 0.01 \rightarrow \Delta AUC_{p^2 = 1} \approx 0.07$) compared to that of the London network ($\Delta AUC_{p^2 = 0} \approx 0.223 \rightarrow \Delta AUC_{p^2 = 1} \approx 0.229$). Additionally, our findings reveal that random data inaccuracy does not always translate to lower percolation effectiveness; bigger inaccuracies do not necessarily correlate with worse outcomes. 

\subsection{\label{sec:boost_and_improve}Robustness boosting and disruption effectiveness}

\begin{figure}[]
    \centering
    \includegraphics[width=0.9\linewidth]{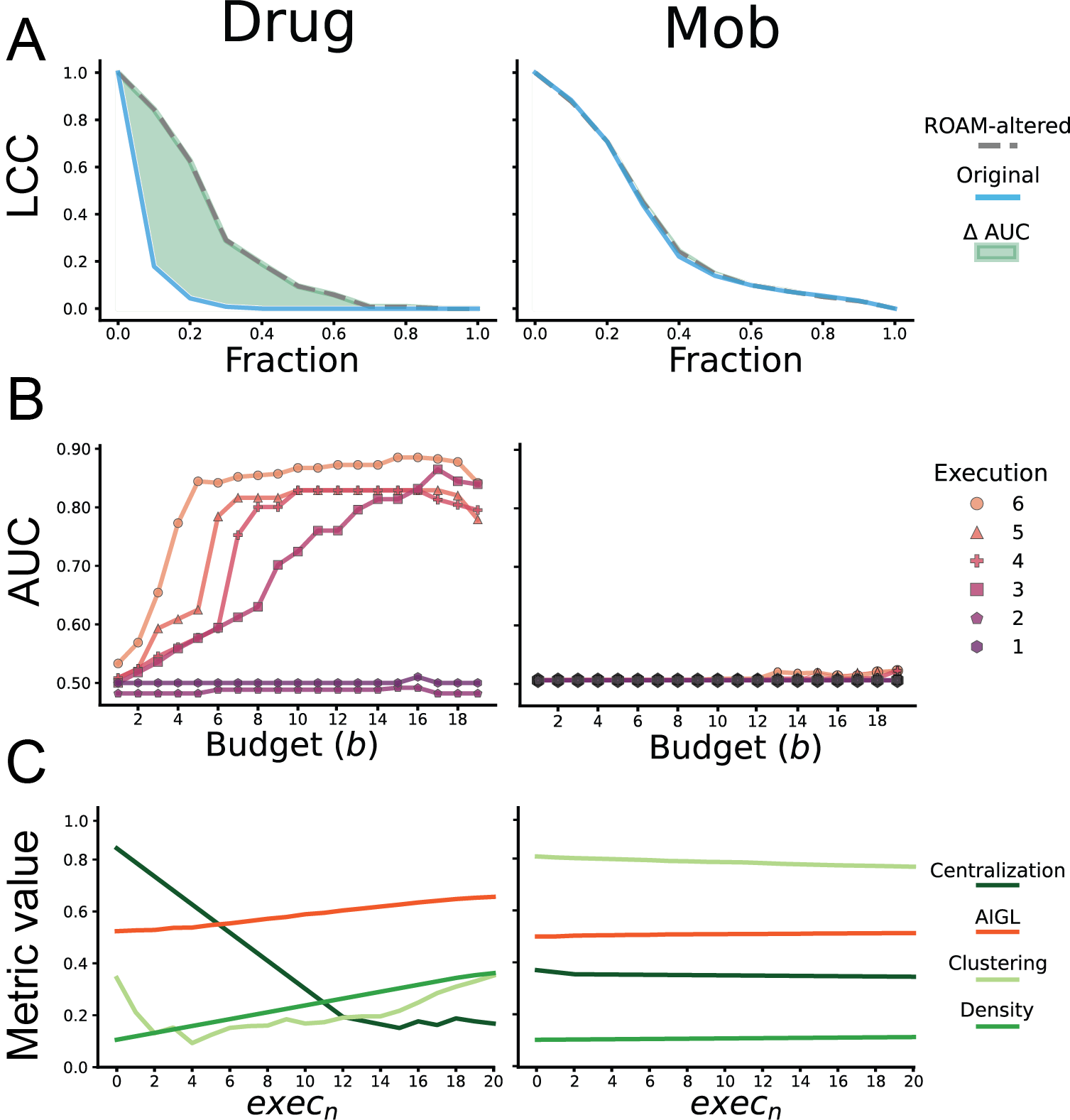}
    \caption{Results of the percolation experiments on the ROAM-altered cocaine trafficking ring and the `Ndrangheta network ($b=6, exec_n = 8$). \textbf{A}: Evolution of LCC under ROAM-altered network (blue line) against the original network (gray dotted line).  Green area indicates the positive difference of AUC between the two trajectories. The trajectories are averaging over all measures as attack strategies. \textbf{B}:  Change in AUC with different values of $b$ and $exec_n$ \textbf{C}: Change in the four network statistics over varying $exec_n$ of ROAM.}
    \label{fig:ROAM_improvement}
\end{figure}

Given the understanding that high centralization reduces network robustness against disruptions, an important question arises: can we enhance network resilience through leader-hiding techniques? In this section, we present the results of numerical experiments aimed at increasing robustness through topological alterations using the ROAM heuristic.

\hyperref[fig:ROAM_improvement]{Figure 3A} shows the results of the percolation experiment on a centralized network (i.e., cocaine) and a decentralized network (i.e., mob) after applying the ROAM heuristic. ROAM starts to be very effective in strengthening a centralized network when $b$ and $exec_n$ are sufficiently large. Particularly, when $b \geq 6$ and $exec_n \geq 8$, the effectiveness of the attack strategies are significantly reduced by a great margin ($0.003 \leq \Delta \text{AUC} \leq 0.15$). Intriguingly, we found that ROAM worked significantly better for the originally centralized networks (i.e., New York cocaine trafficking ring and Madrid bombing network), and was much less effective for networks that are already decentralized. 
While the ROAM-altered centralized networks did not achieve the same level of robustness as the reference baseline (the WS graph), it is reasonable to conclude that ROAM serves not only as a leader-hiding technique to evade detection but also as an effective robustness-enhancing method with a significant impact on the disruption phase of network intervention.

As a next step, we evaluate how the parameterization of the algorithm may affect the effectiveness of such hiding techniques. The parameters quantify  the costs required to run the algorithm in real-life. Particularly, the budget ($b$) and number of executions ($exec_n$) parameters in ROAM were inspected. \hyperref[fig:ROAM_improvement]{Figure 3B} illustrates the gain in robustness as measured by the AUC of various LCC trajectories relative to the original graph in ROAM. Assuming that any modification to edges incurs the same cost, increasing the number of executions almost always yielded better payoff when controlling for the total number of edges that can be modified. For example, under the scenario where the total number of edges modified is 24 (e.g., $b = 4, exec_n = 6$; $b = 6, exec_n = 4$), the payoff is higher when the number of execution is higher than the budget. Therefore, perhaps unsurprisingly, there exists a clear relation effect between budget and number of executions of the algorithm. Nonetheless, it should be emphasized that having a low number of executions inhibits the positive influence of budget on the robustness improvement. A general observation is that having a lower number of executions delays the robustness gain even given the large amount of budget devoted to each execution. Particularly, when $1 \leq exec_n \leq 2$, the network robustness does not improve at all. One significant note to point out is that even with higher number of executions and budget, the positive  effects of ROAM on AUC for the originally decentralized networks become negligible. 

Finally, to understand how ROAM induces structural changes, we conducted additional analyses on the evolution of several global graph properties, namely (1) centralization, (2) average inverse geodesic length (AIGL), (3) average clustering and (4) network density. \hyperref[fig:ROAM_improvement]{Figure 3C} considers the New York cocaine trafficking ring and the `Ndrangheta network. Beyond the two representative networks shown, we observe that across all originally centralized networks, for the networks that gained the starkest increase in AUC, (1) centralization has a negative trend and (2) AIGL, (3) clustering and (4) density have positive trend over the iterations. In other words, via ROAM, the centralized networks are in fact in a process of decentralizing and fostering previously distant connections as observed from the increasing clustering coefficient, AIGL and network density. Nonetheless, echoing our findings earlier in this section, originally decentralized graphs did not experience any visible changes in the structure of the graph (see Mob in \hyperref[fig:ROAM_improvement]{Figure 3C}).

\section{Discussion and conclusion}

Data incompleteness and inaccuracy pose significant challenges to law enforcement agencies (LEAs) in effectively disrupting criminal networks \cite{DeMoor2020} \cite{Ficara2022}. This article aims to explore critical issues in criminal network disruption through the application of network and information-theoretic tools. Our findings reveal that even a 20\% level of missing data can severely impair the effectiveness of leading node-ranking methods. In contrast, data inaccuracy only compromises these methods when a high percentage of actors' information is inaccurately captured. This result highlights that data incompleteness extends beyond simply skewing network statistics; it directly undermines the success of network attack strategies. Specifically, we observed that data incompleteness (i.e., networks with missing nodes) is a more  challenging task than data inaccuracy (i.e., networks with modified edges).

Data incompleteness is a significant threat to both the intelligence and implementation phases of network disruption missions. How, then, can we address this challenge? A critical first step towards a more resilient criminal network disruption strategy is investing in consistent, interoperable data infrastructures for effective data sharing. This goes beyond simply merging as many data sources as possible, as \cite{Duijn2015} suggests; the issue often stems from technical inconsistencies across varied data sources. To maximize both the scope and accuracy of network data, it is essential to synthesize ``large volumes of disparate data” \cite{Spyropoulos2023} (p.3) collected from diverse intelligence channels.
A recent model of such an integrated security system is the EU interoperability regime established under Regulation (EU) 2019/818, which facilitates police and judicial cooperation via systems like the European Criminal Records Information System (ECRIS), the Europol system, and the Prüm II framework \cite{giannakoula_combating_2020}. Prüm II, in particular, aims to ``improve, facilitate and accelerate data exchange” by enabling more open sharing of biometric and criminal records across EU member states \cite{european_commission_new_nodate}.
However, these regulations and systems often lack the formalization necessary for practical network data representation. Interoperability efforts tend to focus on data storage and exchange (e.g., using the universal message format, UMF), yet they neglect crucial aspects of data fusion, such as defining legitimate criminal contacts (e.g., is co-arrest sufficient to establish a link?) and standardizing the labeling of individuals and their relationships. Consequently, constructing a reliable representation of criminal networks remains a challenge for LEAs.
While data-sharing infrastructure continues to evolve, advancements in data collection technology must be matched by collaboration with civil society organizations (e.g., Tech Against Terrorism) and networked investigators (e.g., Bellingcat) to foster a more comprehensive intelligence-gathering process.

Our experiments also reveal that centralized networks are consistently more vulnerable to attacks, though they become more resilient as data quality declines. This finding, combined with highly effective robustness-boosting techniques, is particularly troubling given the trend toward increased decentralization—both technically and socially—in illicit networks over time \cite{McCarthyJones2020}. Notably, the ROAM leader-hiding technique does more than simply obscure key actors; it significantly enhances the resilience of criminal organizations against network attacks, even in relatively centralized networks. This aligns with expectations, as such techniques effectively reduce degree centralization, thus blurring the lines between centralized and decentralized networks.
Consistent with prior research indicating that centralized networks are susceptible to degree-based or value chain-based attacks \cite{Duijn2015}, our results suggest a similar trend: more decentralized networks—both synthetic (e.g., WS, ER) and empirical (e.g., London gangs, `Ndrangheta networks)—tend to be more robust against non-random attack strategies across all evaluation metrics. If criminal networks continue to decentralize in response to technological advances, the detrimental effects of data incompleteness may be even greater than previously estimated.
Furthermore, as this paper focuses on dismantling smaller sub-units of organized crime networks, large-scale dismantling of poly-criminal networks might exhibit different dynamics. According to \cite{european_union_agency_for_law_enforcement_cooperation_decoding_2024}, approximately 20\% of high-risk criminal networks are poly-criminal, meaning they encompass diverse, topologically varied components. With decentralization and poly-criminality on the rise, effectively disrupting these networks remains a critical challenge for LEAs, even with advanced tactics.
For these reasons, alongside developing interoperable intelligence systems, we urge the scientific community to advance rigorous network inference techniques—such as network reconstruction and deep learning approaches—that can accurately integrate information from diverse sources.

In summary, future research could extend this study by examining larger networks, such as cryptocurrency transaction networks and illicit peer-to-peer communication systems, to assess the replicability of our findings. We also encourage the exploration of alternative percolation methods, such as triadic or community-based approaches, which may offer insights into resilience and disruption dynamics in larger and more complex networks. Additionally, percolation experiments on networks with adaptive features—such as recruitment or temporary incarceration modeled through Susceptible-Infected-Recovered (SIR) frameworks—could reveal further realistic impacts of data quality on the effectiveness of intervention strategies.

\section{\label{method}Materials and method}
\subsection{\label{method:data}Data description}
\begin{table}[ht]
\centering
\resizebox{\columnwidth}{!}{%
\begin{tabular}{lcccccc} 
\toprule
\textbf{Type} & \textbf{$N$} & \textbf{$M$} & \textbf{Clustering} & \textbf{Average Distance} & \textbf{Density} & \textbf{Degree centralization} \\ \midrule
\multicolumn{1}{l}{\textbf{Synthetic}} & \multicolumn{6}{c}{} \\ \midrule 
ER & 50 & 149 & 0.09 & 2.36 & 0.12 & 0.11 \\
BAHK & 50 & 140 & 0.52 & 2.32 & 0.11 & 0.39 \\
BA & 50 & 141 & 0.27 & 2.26 & 0.12 & 0.37 \\
WS & 50 & 150 & 0.25 & 2.43 & 0.12 & 0.06 \\
Superlinear DN & 50 & 137 & 0.48 & 3.29 & 0.11 & 0.20 \\ \midrule
\multicolumn{1}{l}{\textbf{Empirical}} & \multicolumn{6}{c}{} \\ \midrule 
London Gangs & 54 & 315 & 0.63 & 2.05 & 0.22 & 0.26 \\
New York Cocaine Trafficking Network & 28 & 40 & 0.34 & 2.07 & 0.11 & 0.84 \\
Ndrangheta mob network & 139 & 1470 & 0.81 & 2.33 & 0.15 & 0.37 \\
Madrid Bombing Terrorist Network & 17 & 63 & 0.90 & 1.59 & 0.46 & 0.54 \\ \bottomrule
\end{tabular}%
}
\caption{Data description of the the networks being investigated. $N$ and $M$ indicate the number and edges of the networks respectively. The synthetic networks are generated with approximately constant density to control for the network connectivity for comparative purposes.}
\label{tab:data-description}
\end{table}

The empirical networks were primarily collected through evidence presented in courts with edges between individuals representing an incidence of communication, including but not limited to tapped phone calls, mob conferences and co-appearance in arrests. To validate our result, we generated various unweighted, undirected graphs with several generative models, namely (1) Erdős–Rényi model (ER); scale-free networks with (2) Barabási-Albert model (BA) \cite{Albert2002} and (3) Holme and Kim’s variation of the BA model (BAHK) \cite{HolmeKim2002}; (4) small-world network with Watts-Strogatz model (WS) \cite{Watts1998}; (5) superlinear densifying network (SDN) \cite{Lambiotte2016}. The models were parameterized to minimize the difference in network density and number of edges to control for the connectivity in the graphs.

\subsection{\label{method:experiment}Numerical experiment}

\subsection{\label{method:startegy}Attack strategies}
The following sections will detail the mathematical rationale of the node ranking tactics. Note that we do not consider bond percolation in this work as site percolation is considerably more efficient in graph dismantling tasks \cite{Holme2002}.

\begin{table}[ht]
\resizebox{\columnwidth}{!}{%
\begin{tabular}{@{}lll@{}}
\toprule
\textbf{Category}         & \textbf{Method}           & \textbf{Computational Complexity} \\ \midrule
\textbf{Centrality-based} & Degree centrality         & $O(m)$                            \\
                          & Eigenvector centrality    & $O(km)$                           \\
                          & Katz centrality           & $O(n^3)$                          \\
                          & PageRank centrality       & $O(km)$                           \\
                          & Closeness centrality      & $O(nm)$                           \\
\textbf{}                 & Betweenness centrality    & $O(nm)$                           \\
\textbf{Heuristics-based} & Collective Influence (CI) & $O(n \log n)$                     \\
\textbf{}                 & Core High-Degree (CoreHD) & $O(n)$                            \\
\textbf{Machine Learning} & FINDER                    & $O(n+m+n \log n)$                 \\
\bottomrule
\end{tabular}%
}
\caption{Node-ranking methods used in this work. Computational complexity is assumed optimal for sparse networks. Note that for eigenvector and PageRank, $k$ indicates the iterations needed for convergence.}
\label{tab:attack-metrics-table}
\end{table}

\begin{figure}[ht]
    \centering
    \includegraphics[width=\linewidth]{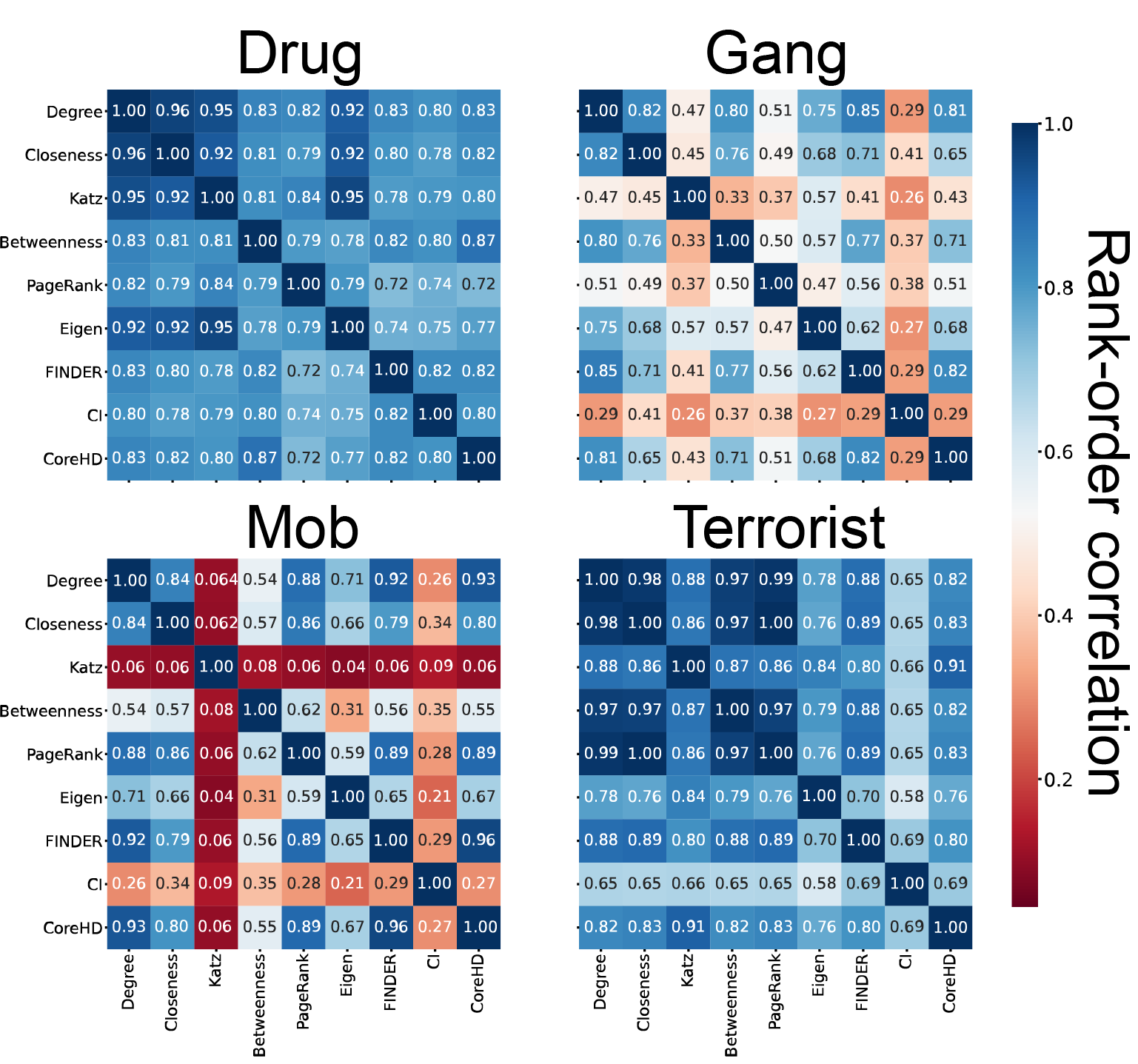}
    \caption{Rank-biased overlap similarity of node-ranking methods used in the percolation experiment. Higher values indicate higher similarity between node-ranking methods.}
    \label{fig:fig4}
\end{figure}

\subsubsection{\label{method:heuristics}Heuristics-based attacks}
Other than the classical centrality measures (see \autoref{tab:attack-metrics-table}), we also used two heuristics-based method for targeting. CI is a heuristic that search for the minimal set of influencers to be targeted to reduce their influence in a network as seen in a typical influence maximization problem \cite{Morone2015}. CI index of a node given by

\begin{equation}
    \text{CI}(v_i) = (k_i - 1) \sum_{j \in \partial B(i, \ell)} (k_j - 1),
\end{equation}
where $k_i$ is the degree of node $v_i$, $B(i, \ell)$ is the ball centring on node $v_i$ and $\partial B(i, \ell)$ are the nodes at the frontiers of the ball. In simple terms, CI of a length $\ell = 2$ is the product between the sum of the degree of all nodes located at the shortest path distance exactly at 2 from node $v_i$ and the degree of the node $v_i$ itself. This quantity is a scalable method to search for  minimal sets of nodes \cite{Morone2015}, which is extremely useful for finding influential players in very large complex networks. \cite{Morone2016} developed an even more efficient computational method using max-heap, wuth the computational complexity $O(n\log n)$. 

Another heuristic is Core High-Degree (CoreHD), an approach that utilizes degree-based decycling - the disintegration of cycles \cite{Zdeborov2016}. The algorithm of CoreHD is rather simple: 

\begin{algorithm}
\caption{CoreHD}
\begin{algorithmic}[1]
\STATE Find the kcore and obtain the degree of the nodes within the kcore
\WHILE{$|\text{kcore}| > 0$}
\STATE Find the set of nodes $V_{\text{HD}}$ with the highest degree
\IF{$|V_{\text{HD}}| > 1$}
    \STATE Randomly choose one
\ENDIF
\STATE Remove the chosen node, update the kcore and the degrees of the nodes 
\ENDWHILE
\STATE Tree-breaking and greedy reinsertion
\end{algorithmic}
\end{algorithm}

This method is similar to the degree centrality-based attacks, but it focuses on adaptive node removal inside the kcore with greedy reinsertion, a technique designed to minimize the number of unnecessary nodes removed during the process of decycling. This method is extremely fast with a computational complexity of $O(n)$ for sparse networks and generally as effective as other message-passing decycling attacks such as Min-Sum. 

\subsubsection{\label{method:ml}Machine Learning-based attacks}
We also consider a popular pre-trained Graph Neural Network model (GNN), the so-called FInding key players in Networks through DEep Reinforcement learning (FINDER) \cite{Fan2020}, in order to target nodes in the network. FINDER is a deep reinforcement learning framework for optimal percolation problems proposed by \cite{Fan2020}. Formally, FINDER aims to minimize the accumulated normalized connectivity (ANC)

\begin{equation}
    R(v_1, v_2, \cdots, v_N) = \frac{1}{N} \sum_{k = 1}^N \frac{\sigma (G \symbol{92} \{v_1, v_2, \cdots, v_k\})}{\sigma (G)},
\end{equation}
where $N$ is the the number of nodes in $G$ and $v_i \in V$ indicates the $i$th node to be percolated from the graph, $\sigma (G \symbol{92} \{v_1, v_2, \cdots, v_k\})$ is the connectivity of the graph after $v_i$ is removed from the initial graph and, intuitively, $\sigma (G)$ is the connectivity of the original graph. The connectivity metric in FINDER can be any well-defined network metric (i.e., network robustness metrics), making it an extremely adaptable method to consider different types of metrics, such as Von Neumann entropy and spectral gap. 

The model contains two phases, namely the offline training phase and the online application phase. In the first phase, synthetic graphs are generated based on different network generative models. These graphs are randomly sampled for the agent to play the game - an episode of a crucial node identification process - where the agent's action is to remove the chosen node. As mentioned, the reward to such an action is defined by the ANC, and the larger the marginal decrease of ANC, the more reward an agent will obtain. The graphs are encoded with tunable parameters $\Theta_e$ using inductive graph representational learning to aggregate node embedding vectors as node features to obtain their latent structural position in a graph. After capturing the node embedding, the embedding is then decoded with tunable parameters $\Theta_d$ as a scalar $Q$, a set of scores that assesses the potentials of any given actions. A multilayer perceptron with RELU activation is used to generate the output layer containing the $Q$ values. Using the $\epsilon$-greedy strategy under an exploration-exploitation framework, the action with the highest $Q$ will be adopted with a probability of (1- $\epsilon$), and a random action will be undertaken with a probability of $\epsilon$. $\epsilon$ decreases linearly from 1.0 to 0.05 over episodes, symbolizing that a more experienced agent will make decisions based on its past learning (i.e., exploitation) rather than exploring new options in comparison to a less experienced agent. When a game is completed, $n$-step transitions $(S_i, A_i, R_{(i, i+n)}, S_{i+n})$ are collected. $M$ most recent transitions are then stored in the experience replay buffer, a memory storage technique commonly used Deep-Q learning models. With these memories, the agent is updated with a new set of parameters $\Theta_e$ and $\Theta_d$ for the encoder and decoding processes respectively. Adam gradient descent updates are used to compute the loss from the randomly sampled experiences from the $M$ most recent memories. 

During the application phase, the empirical network will be fed to the model and encoded into a embedding vector with lower dimensions, and then the model will infer the $Q$-value for each node (i.e., the reward when such node is removed from the network) using the decoder. Note that in practice the model selects nodes in batches that can maximize $Q$ instead of the computing $Q$ for each node to reduce computational complexity to $O(|E| + |V| + |V|\log |V|)$. For example, instead of choosing the node $v_0$, FINDER will evaluate the $Q$-value of removing the set of $\{v_0, v_1, v_2, v_5\}$.

The model used in this paper is pre-trained by the authors in the original FINDER paper. The model was trained on ER, WS and BA networks ($n_{\text{total}} = 2 \times 10^6$) with each synthetic network containing 30 to 50 nodes. The model sets $M = 5 \times 10^4$ for the experience replay buffer. Because of the batch selection strategy, FINDER  only returns the most effective set of nodes to be percolated from the complete graph, meaning that some nodes will be omitted from FINDER. To resolve this problem, the final ordered list of nodes to be removed from graph contains two separate subsets: (1) one with the nodes provided by FINDER and (2) one imputed randomly with the residual nodes in the network. Because subset 1 is not ordered, it is possible that the effectiveness of FINDER under our scenario of sequential node removal is not optimal. However, FINDER has overall been proven to be effective even compared to node rankings that are theoretically designed to be completely ordered. 

\subsection{\label{method:ROAM}Remove-One-Attach-Many heuristic}

\begin{figure}[ht]
\centering
\includegraphics[width=\textwidth]{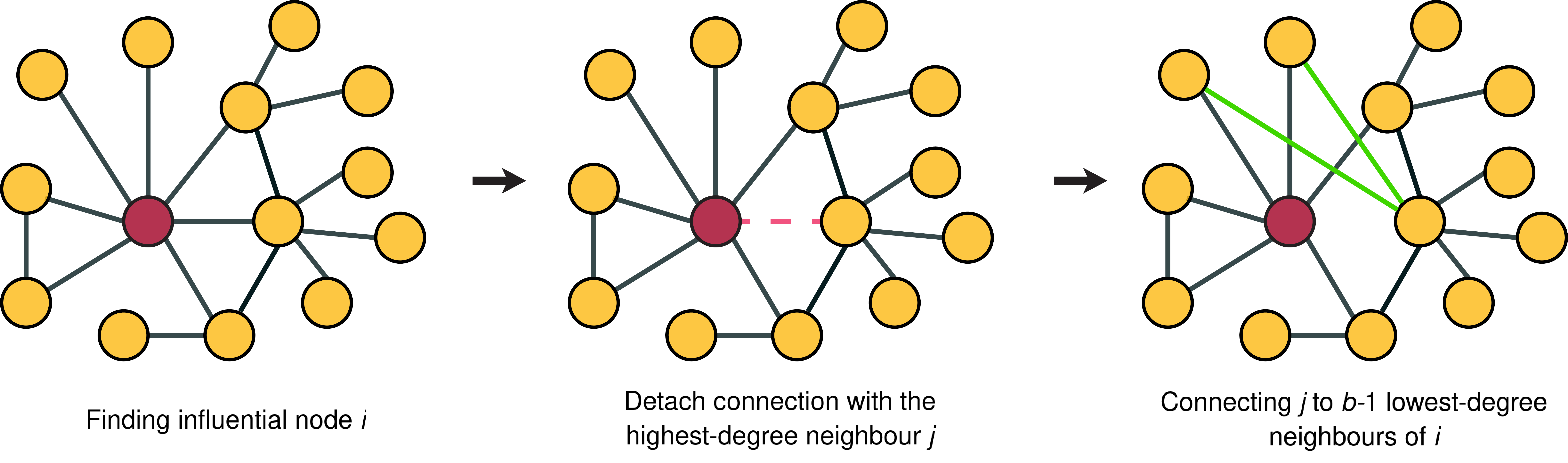}
\caption{One execution of the Remove-One-Attach-Many heuristic with a budget $b = 3$.}
\label{ROAM_schematic} 
\end{figure}
Finally, we also used a leader-hiding heuristic called Remove-One-Attach-Many (ROAM), an efficient method to hide the leading actor in a network simply by rewiring links of the most central person in a network (see \autoref{ROAM_schematic}) \cite{waniek_hiding_2018}, to boost the resilience/robustness of a network. The algorithm takes two parameters, $b$ and $exec_n$, which are the budget available for link addition and removal (see \autoref{ROAM_schematic}) and the number of consecutive execution of ROAM respectively. In practice, ROAM searches for the evader $v^\dagger$. It then detaches the edge between $v^\dagger$ and its most connected neighbour $v_0$. By doing so, we reduce the centrality of the leader. Then, to recover the loss of influence of the leader due to reduced connections, we artificially add $b-1$ links between $v_0$ and its least connected neighbours. The success of ROAM, then, relies on $b$ as well as how many times we execute the heuristic. In this work, the evader was chosen to be the actor with the lowest combined rank of degree, betweenness and closeness centrality measures, resembling a leader in the network.

\backmatter

\section*{Declarations}

\subsection*{Ethics approval and consent to participate}
Not applicable
\subsection*{Consent for publication}
Not applicable
\subsection*{Availability of data and material}
The code will be made available after publication. The datasets for the replication of this work are publicly available and can be found via the following links: 

\begin{enumerate}
    \item \href{https://sites.google.com/site/ucinetsoftware/datasets/covert-networks/london-gang}{Gang network}
    \item \href{https://comeetie.github.io/greed/reference/Ndrangheta.html}{Mob Network}
    \item \href{https://sites.google.com/site/ucinetsoftware/datasets/covert-networks/cocaine-dealing-natarajan}{Drug trafficking network} 
    \item \href{https://networks.skewed.de/net/train_terrorists}{Terrorist network}  
\end{enumerate}

\subsection*{Competing interests}
The authors declare that they have no competing interests.
\subsection*{Funding}
R.L. acknowledges support from the EPSRC grants EP/V013068/1, EP/V03474X/1 and EP/Y028872/1. 
\subsubsection*{Author contribution}
W.N.Y and R.L. conceived the study, developed the theoretical framework and designed the experiments. W.N.Y. performed all experiments, computation and analyses. R.L. supervised the project. W.N.Y. and R.D.C. produced the figures. All authors discussed the results and contributed to the final manuscript.
\subsection*{Acknowledgments}
The authors thank Bernie Hogan (Oxford Internet Institute) for his ideas during the conception phase of this work and Rafael Prieto-Curiel (Complexity Science Hub) for his insightful feedback on this work during and after the Conference on Complex Systems `24 in Exeter, United Kingdom.

\newpage
\bibliography{sn-bibliography}

\end{document}